\documentclass[pdflatex,sn-aps]{sn-jnl}
\usepackage{graphicx}%
\usepackage{multirow}%
\usepackage{amsmath,amssymb,amsfonts}%
\usepackage{amsthm}%
\usepackage{mathrsfs}%
\usepackage[title]{appendix}%
\usepackage{xcolor}%
\usepackage{textcomp}%
\usepackage{manyfoot}%
\usepackage{booktabs}%
\usepackage{algorithm}%
\usepackage{algorithmicx}%
\usepackage{algpseudocode}%
\usepackage{listings}%
\usepackage{dcolumn}
\usepackage{bm}
\usepackage{braket}
\usepackage{color}
\usepackage{accents}
\usepackage{mathrsfs}
\usepackage{bm}
\usepackage{here}
\usepackage{longtable}

\usepackage{mathtools}
\usepackage{empheq}
\usepackage{comment}
\usepackage{float}

\theoremstyle{thmstyleone}%
\theoremstyle{thmstyletwo}%

\theoremstyle{thmstylethree}%

\begin{document}

\title[]{Orbital magnetic octupole in crystalline solids and characterization of orbital altermagnetism}

\author*{\fnm{Takumi} \sur{Sato}}\email{sato@phys.sci.hokudai.ac.jp}
\author{\fnm{Satoru} \sur{Hayami}}\email{hayami@phys.sci.hokudai.ac.jp}
\affil{\orgdiv{Graduate School of Science}, \orgname{Hokkaido University}, 
\orgaddress{\city{Sapporo}, \postcode{060-0810}, \country{Japan}}}

\abstract{
Magnetic multipole moments beyond dipoles have emerged as key descriptors of unconventional electromagnetic responses in crystalline solids. 
However, a gauge-invariant bulk expression for orbital magnetic multipole moments has remained elusive, hindering a unified understanding of their physical consequences.
Here we formulate a gauge-invariant expression for the orbital magnetic octupole moment in periodic crystals and investigate its behavior 
in two models with distinct origins of magnetism: 
a spinless two-orbital model with orbital magnetic order arising from complex hopping and 
a two-sublattice $d$-wave altermagnetic model based on antiferromagnetic spin order. 
We also show that the orbital magnetic octupole is naturally linked to a higher-rank
Hall response induced by spatially nonuniform electric fields, leading to a generalized St\v{r}eda-type relation. 
Our results further demonstrate that the orbital magnetic octupole provides 
a quantitative characterization of \textsl{orbital altermagnetism}. 
}




\maketitle

\section{Introduction}
\label{sec:intro}
In the active search for novel magnetic materials, a central challenge
is to efficiently and comprehensively characterize nontrivial electronic states 
beyond conventional ferromagnets (FMs) and antiferromagnets (AFMs).
In this context, magnetic multipoles have emerged as powerful descriptors for capturing 
essential features of electronic states in magnetic systems, including unconventional transport and magnetoelectric responses%
~\cite{spaldin:jpcm2008, kusunose:jpsj2008, kuramoto:ptp2008review, kuramoto:jpsj2009review, santini:rmp2009,
watanabe:prb2017, suzuki:jpsj2018review, hayami:prb2018classification, watanabe:prb2018classification, yatsushiro:prb2021classification, 
hayami:jpscp2020, hayami:jpsj2019spinsplit, hayami:prb2020antisymspinsplit, hayami:prb2020spinsplit,
hayami:jpsj2024review, watanabe:jpcm2024review}.
Typically, the magnetic dipole (MD) moment is finite and provides a quantitative description of magnetization in FMs.
Interestingly, altermagnets (AMs) constitute an unconventional class of collinear antiferromagnets that
exhibits momentum-dependent spin-split band structures and broken time-reversal symmetry while preserving vanishing net magnetization%
~\cite{noda:pccp2016, okugawa:jpcm2018, smejkal:sciadv2020, naka:natcomm2019, ahn:prb2019, 
hayami:jpsj2019spinsplit, hayami:jpscp2020, hayami:prb2020spinsplit, yuan:prb2020, naka:prb2021perovskite, 
yuan:prm2021, yuan:prb2021, hernandez:prl2021, smejkal:prx2022magnetoresistance, mazin:pnas2021, 
smejkal:prx2022spingroup1, smejkal:prx2022spingroup2, 
cheong:npj2025AMclassification, guo:advmat2025review, hu:prx2025}.
As a result, AMs do not possess an MD moment that couples directly to a uniform magnetic field.
Instead, they can host higher-rank magnetic multipole moments, most notably a magnetic octupole (MO), which
couples to spatially varying magnetic fields~\cite{spaldin:jpcm2008,bhowal:prx2024}.

In classical electromagnetism, the MO $M_{ijk}$ is defined by~\cite{jackson}
\begin{align}
    M_{ijk} = \frac{1}{4cV_{\mathrm{vol}}} \int d\bm{r}\, [ \bm{r} \times \bm{j} ]_i r_j r_k 
    \label{eq:MO_classical},
\end{align}
where $\bm{j}$ and $\bm{r}$ are the current density and the position, respectively.
However, Eq.~\eqref{eq:MO_classical} cannot be directly applied to periodic crystals 
because the position operator is ill-defined in the Bloch representation. 
While similar difficulties arise for other electric and magnetic multipole moments, 
and have been successfully resolved by the modern theory of multipole moments%
~\cite{king-smith:prb1993, vanderbilt:prb1993, resta:rmp1994, resta:book, resta:jpcm2010review, xiao:rmp2010, vanderbilt:book, 
resta:chemphyschem2005, xiao:prl2005, xiao:prl2006thermoele, thonhauser:prl2005, ceresoli:prb2006, 
shi:prl2007, souza:prb2008, thonhauser:ijmpb2011, 
bianco:prl2013,seleznev:prb2023,
gao:prb2018spinMQM, shitade:prb2018orbitalMQM, gao:prb2018orbitalMQM, shitade:prb2019spinMQM, 
daido:prb2020, 
oike:prb2025spinMO, TS:npjqmat2026},
a corresponding microscopic formulation of the orbital MO moment for crystalline solids has remained missing.

In this paper, we establish a microscopic theory of the orbital MO in crystalline solids based on quantum mechanics and thermodynamics%
~\cite{shi:prl2007, gao:prb2018spinMQM, shitade:prb2018orbitalMQM,
gao:prb2018orbitalMQM, shitade:prb2019spinMQM, daido:prb2020, oike:prb2025spinMO, TS:npjqmat2026}. 
By evaluating the free-energy density up to the second-order spatial derivative of the magnetic field, 
we derive a gauge-invariant bulk expression for the orbital MO. 
While the spin contribution to the MO has been recently formulated~\cite{oike:prb2025spinMO,TS:npjqmat2026}, 
the present study provides the previously missing orbital contribution.

We further clarify the thermodynamic relation connecting the orbital MO to physical response functions. 
In particular, we show that the orbital MO is directly related to the quadrupolar magnetoelectric (QME) effect, which 
describe electric polarization (magnetization) induced by magnetic (electric) field gradients%
~\cite{oike:prb2025spinMO, TS:npjqmat2026, shitade:prb2025}.
Moreover, we demonstrate that the QME effect is intimately connected to the octupolar anomalous Hall (OAH) effect, 
which characterizes the anomalous Hall response driven by a nonuniform electric field~\cite{kozii:prl2021,TS:npjqmat2026}, 
as a higher-rank analogue of the St\v{r}eda formula~\cite{streda:jphysc1982,widom:physletta1982}.

Based on the established formulation, we apply the orbital MO formula 
to two representative models with distinct microscopic origins of time-reversal-symmetry breaking: 
a spinless two-orbital model with orbital magnetic order and 
a two-sublattice AFM with $d$-wave spin splitting. 
In the former model, time-reversal symmetry is broken by complex hopping, 
whereas in the latter it is broken by antiferromagnetic molecular fields. 
We demonstrate that relativistic spin-orbit coupling (SOC) is necessary for a finite orbital MO in the AFM model, 
while the spinless two-orbital model does not require SOC. 
In both models, we further show that the OAH effect can remain finite 
even when the conventional linear anomalous Hall response to a uniform electric field is symmetry-forbidden, 
in which case it constitutes the leading-order Hall response. 
We also find that the orbital MO can quantify \textsl{orbital altermagnetism}%
~\cite{yu:natcomm2025,sobral:prr2025,pupim:prl2025,chakraborty:arxiv2025orbitalAM,pan:arxiv2025orbitalAM}, 
just as the spin MO quantifies ordinary spin altermagnetism~\cite{oike:prb2025spinMO,TS:npjqmat2026}. 
Our results provide a framework for exploring orbital MO-ordered systems beyond symmetry-based considerations.
Throughout this paper, we use the units of $k_\mathrm{B} = c = \hbar = 1$
where $k_\mathrm{B}$ is the Boltzmann constant and $c$ is the speed of light
and $e<0$ is the charge of an electron.

\section{Results and Discussion}
\label{sec:results}
\subsection{Orbital MO in crystalline solids}
\label{subsec:MO_formulation}
To formulate the orbital MO
in periodic systems without explicitly invoking the ill-defined position operator, we adopt a thermodynamic approach based on the free energy.
We consider a perfectly periodic crystal subjected to a slowly varying magnetic field 
$\bm{B}(\bm{r}) = \nabla \times \bm{A}(\bm{r})$, where $\bm{A}(\bm{r})$ is a vector potential.
In such a situation, the differential form of the free-energy density $F(\bm{r})$ is described as%
~\cite{shi:prl2007,gao:prb2018spinMQM,shitade:prb2018orbitalMQM,gao:prb2018orbitalMQM,shitade:prb2019spinMQM,oike:prb2025spinMO,TS:npjqmat2026}
\begin{align}
    dF =
    &-SdT - Nd\mu - M_{i} dB_{i}
    -M_{ij} d[\partial_{r_j} B_{i}] - M_{ijk} d[\partial_{r_j} \partial_{r_k} B_{i}] 
    \label{eq:differential_FE} ,
\end{align}
where $S$, $T$, $N$, and $\mu$ are the entropy, temperature, particle number, and chemical potential, respectively.
The orbital magnetic multipoles are defined as the quantity conjugate to 
the magnetic field and its spatial derivatives in Eq.~\eqref{eq:differential_FE}, 
where $M_i$, $M_{ij}$, and $M_{ijk}$ ($i,j,k = x, y, z$) denote 
the orbital MD, orbital magnetic quadrupole, and orbital MO, respectively.
Accordingly, the orbital MO in periodic crystals is described as
\begin{align}
    M_{ijk} \coloneqq - \left( \frac{\partial F}{\partial[\partial_{r_j} \partial_{r_k} B_i]} \right)_{T,\mu,\bm{B},\partial_{\bm{r}}\bm{B}} .
    \label{eq:def_MO}
\end{align}
For the purpose of the derivation of the orbital MO, we first calculate the particle number change 
due to the second-order spatial derivative of the magnetic field $\partial_{r_j} \partial_{r_k} B_{i}$.
This strategy exploits the follwing Maxwell relation~\cite{oike:prb2025spinMO}: 
\begin{align}
    \frac{\partial M_{ijk}}{\partial \mu} &= \frac{\partial N}{\partial[\partial_{r_j} \partial_{r_k} B_i]} ,
    \label{eq:maxwell}
\end{align}
which follows directly from Eq.~(\ref{eq:differential_FE}).
The orbital MO is then obtained by integrating Eq.~\eqref{eq:maxwell} with respect to $\mu$. 
Compared with a direct evaluation of Eq.~\eqref{eq:def_MO}~\cite{shitade:prb2019magnonGME,daido:prb2020,TS:npjqmat2026}, 
this approach simplifies analytical calculations~\cite{oike:prb2025spinMO}.

To obtain $\partial N / \partial[\partial_{r_j} \partial_{r_k} B_i]$, we expand 
the density-current correlation function,
\begin{align}
    \chi_{N, J_a}(\bm{q},\omega) = 
    -\sum_{nm} \int \frac{d^d k}{(2\pi)^d}
    &\Braket{u_{n\bm{k}_{-}}|u_{m\bm{k}_{+}}}
    \Braket{u_{m\bm{k}_{+}}|\frac{e}{2} \left[ \hat{v}_{a}(\bm{k}_{+}) + \hat{v}_{a}(\bm{k}_{-}) \right]|u_{n\bm{k}_{-}}} \nonumber \\
    \times \, &\frac{f(\epsilon_{n\bm{k}_{-}}-\mu)-f(\epsilon_{m\bm{k}_{+}}-\mu)}{\omega+i\delta + \epsilon_{n\bm{k}_{-}} -\epsilon_{m\bm{k}_{+}}} 
    \label{eq:chi_nj} ,
\end{align}
which describes the linear response $\delta \langle N \rangle (\bm{q},\omega) = \chi_{N, J_a} (\bm{q},\omega) A_a (\bm{q},\omega)$.
We expand this correlation function up to the third order in the wave vector $\bm{q}$.
Here, the following notations are used:
\begin{align}
    &\hat{H} \Ket{\psi_{n\bm{k}}} = \epsilon_{n\bm{k}} \Ket{\psi_{n\bm{k}}} , \quad
    \Ket{u_{n\bm{k}}} = e^{-i\bm{k}\cdot \hat{\bm{r}}} \Ket{\psi_{n\bm{k}}} , \\
    &\hat{v}_a (\bm{k}) = \partial_{k_a} \hat{H}(\bm{k}) , \quad
    \bm{k}_{\pm} = \bm{k} \pm \frac{\bm{q}}{2} , \quad
    f(z) = [e^{\beta z}+1]^{-1} ,
    \label{eq:notations}
\end{align}
where $\hat{H}$ is the single-electron Hamiltonian in a periodic crystal and 
$\hat{H} ({\bm{k}}) = e^{-i\bm{k}\cdot \hat{\bm{r}}} \hat{H} e^{i\bm{k}\cdot \hat{\bm{r}}}$ is the Bloch Hamiltonian.

We calculate the orbital MO $M_{ijk}$ according to
\begin{align}
    M_{ijk} &= \int_{-\infty}^{\mu} d\mu' \, \frac{\partial M_{ijk}(\mu')}{\partial \mu'} , \\
    \frac{\partial M_{ijk}}{\partial \mu} &= -\frac{i}{12} \epsilon_{iab} \lim_{\bm{q} \rightarrow \bm{0}}
    \partial_{q_b}\partial_{q_j}\partial_{q_k} \lim_{\delta \rightarrow +0} \chi_{N, J_a} (\bm{q},0) ,
\end{align}
where $\epsilon_{iab}$ is the totally antisymmetric tensor.
After performing the algebra, we obtain the expression for the orbital MO in crystalline solids:
\begin{align}
    M_{ijk} &= \frac{e}{12} \sum_{n} \int \frac{d^d k}{(2\pi)^d} 
    \left( A^{ijk}_n f'_n + B^{ijk}_n f_n + C^{ijk}_n \mathcal{G}_n \right) 
    \label{eq:Mijk} ,\\
    A^{ijk}_n &= \epsilon_{iab} 
    \left( \frac{1}{24} m_n^{ab} \partial_{j} \partial_{k} \epsilon_n 
    + (b \leftrightarrow j \leftrightarrow k) \right) 
    \label{eq:Aijk} , \\
    B^{ijk}_n &= \epsilon_{iab} 
    \Bigg( -\frac{1}{24} \partial_{j} \partial_{k} m_n^{ab} 
    - \frac{1}{16} \Omega_n^{ab} \partial_{k_j} \partial_{k_k} \epsilon_n 
    + \frac{3}{4} m_n^{ab} g_n^{jk} \nonumber \\
    &- \mathrm{Im} \Big[ \frac{1}{12} \Braket{D_a n| \hat{v}_{bjk}|n} + \frac{1}{8} \Braket{D_a n| \hat{v}_{jk}|D_b n} \Big]
    + \sum_m^{\neq n} \bigg\{ 
        \frac{1}{8} \frac{\Omega_{nm}^{ab}}{\epsilon_{nm}} v^j_n ( 3v^k_n + 2v^k_m ) \nonumber \\
        &+ \frac{1}{8 \epsilon_{nm}} \mathrm{Im} \big[ \Braket{D_b n | (\hat{v}_a - v^a_n)|m} v^{jk}_{mn} 
        + 2 \Braket{D_b n|\hat{v}_a|m} \Braket{m|\hat{v}_j|D_k n} - (a \leftrightarrow b) \big] \nonumber \\
        &-\frac{1}{2\epsilon_{nm}} \mathrm{Im} \big[ \Braket{n|D_a m} \Braket{m|\hat{v}_b|D_j n}
        + \Braket{n|D_b m} \Braket{D_j m|\hat{v}_a| n} \big] v^k_n \nonumber \\
        &+\frac{1}{2} \mathrm{Im} \big[ \Braket{n|D_a m} \Braket{D_b m | D_j n} \big] v^k_n
    \bigg\} + (b \leftrightarrow j \leftrightarrow k) \Bigg) 
    \label{eq:Bijk} , \\
    C^{ijk}_n &= \epsilon_{iab} \Bigg( -\frac{1}{8} \partial_j \partial_k \Omega_n^{ab} + 
        \sum_m^{\neq n} \Bigg\{ 
            \frac{1}{4} \Omega_{nm}^{ab} \bigg( \frac{\partial_j \partial_k \epsilon_{nm}}{\epsilon_{nm}}
            + g_n^{jk} + g_m^{jk} \nonumber \\
            &-\frac{1}{\epsilon_{nm}^2} \{v^j_n v^k_n + v^j_m v^k_m + \frac{1}{2}(v^j_n+v^j_m)(v^k_n+v^k_m)\} \bigg) \nonumber \\
            &+\mathrm{Im} \bigg[ \Braket{n|D_a m} \Big\{
            \big\{ \Braket{D_j m |\hat{v}_b | n} - \Braket{m|\hat{v}_b|D_j n} \big\}  \frac{-1}{2} \frac{v^k_n+v^k_m}{\epsilon_{nm}^2} \nonumber \\
            &+\big\{ v^b_n \Braket{D_j m|\hat{v}_k|n} - v^b_m \Braket{m|\hat{v}_k|D_j n} \big\} \frac{1}{2 \epsilon_{nm}^2} 
            + \Braket{D_j m| \hat{v}_b|D_k n} \frac{1}{2 \epsilon_{nm}} \nonumber \\
            &+(v^b_n - v^b_m) v^{jk}_{mn} \frac{1}{4\epsilon_{nm}^2} 
            - (a \leftrightarrow b)    \Big\} \bigg] \nonumber \\
            &-\frac{3}{4} \mathrm{Im} \bigg[ \Braket{n|D_a m} \Braket{D_b m|D_j n} \bigg] \frac{v^k_n+v^k_m}{\epsilon_{nm}} \nonumber \\
            &+\frac{1}{4} \mathrm{Im} \bigg[ \Braket{n|D_a m} \Big\{ \Braket{D_j m|\hat{v}_b|n}-\Braket{m|\hat{v}_b|D_j n}\Big\} \bigg] \frac{v^k_n+v^k_m}{\epsilon_{nm}^2} \nonumber \\
            &-\frac{1}{2\epsilon_{nm}} \mathrm{Im} \bigg[ \Big\{ \Braket{D_b m|\hat{v}_a|n} - \Braket{m|\hat{v}_a|D_b n} \Big\} \Braket{D_j n |D_k m} \bigg] \nonumber \\
            &-\frac{1}{6\epsilon_{nm}} \mathrm{Im} \bigg[ \Braket{n|D_a m} v^{bjk}_{mn} \bigg] \nonumber \\
            &+\sum_{\ell}^{\neq n,m} \frac{1}{4\epsilon_{nm}} \bigg\{ \frac{1}{\epsilon_{n\ell}} \mathrm{Im} \bigg[
                \Big\{ v^{jk}_{\ell n} + 2\Braket{\ell|(\hat{v}_j - v^j_n)|D_k n} \Big\} \Braket{n|D_b m} v^a_{m\ell} - (a \rightarrow b)
                \bigg] \nonumber\\
                &+ (n \leftrightarrow m) \bigg\}
        \Bigg\} + (b \leftrightarrow j \leftrightarrow k) \Bigg) 
        \label{eq:Cijk} ,
\end{align}
where we used the following abbreviations:
$\ket{n}=\ket{u_{n\bm{k}}}$, $\partial_a=\partial_{k_a}$, 
$\hat{v}_a = \hat{v}_a(\bm{k})$, $\hat{v}_{ab} = \partial_b \hat{v}_a$, $\hat{v}_{abc} = \partial_c \hat{v}_{ab}$,
$v^{a\cdots}_{nm}=\Braket{n|\hat{v}_{a\cdots}|m}$, $v^{a\cdots}_{n}=v^{a\cdots}_{nn}$,
$\epsilon_n = \epsilon_{n\bm{k}}$, $\epsilon_{nm} = \epsilon_n - \epsilon_m$,
$\mathcal{G}_n = -T\log (1+e^{-(\epsilon_n-\mu)/T})$, 
$f_n = f(\epsilon_n-\mu)$, $f'_n = \partial f(z) / \partial z|_{z=\epsilon_n-\mu}$.
Here, the symbol $(b \leftrightarrow j \leftrightarrow k)$ denotes taking the sum over all expressions generated by permutations of the indices:
$F^{abjk} + (b \leftrightarrow j \leftrightarrow k) =
F^{abjk} + F^{akbj} + F^{ajkb} +
F^{abkj} + F^{akjb} + F^{ajbk}$.
$D_a$ is the covariant derivative acting to the Bloch state, defined as 
$\ket{D_a n} = \hat{Q}_n \ket{\partial_a n}$ with $\hat{Q}_n=1-\ket{n}\bra{n}$.
Under the gauge transformation $\ket{n} \rightarrow e^{i\eta_n} \ket{n}$, 
$\ket{D_a n}$ transforms as $\ket{D_a n} \rightarrow e^{i\eta_n} \ket{D_a n}$ ($\eta_n=\eta_n(\bm{k})\in \mathbb{R}$), 
which guarantees the gauge invariance of Eq.~\eqref{eq:Mijk}.
In Eqs.~\eqref{eq:Aijk}-\eqref{eq:Cijk}, we also defined the following gauge invariant quantities:
\begin{align}
    m_n^{ab}     &=  -\mathrm{Im}\, \tilde{T}_n^{ab} ,\quad
    \Omega_n^{ab} = -2\mathrm{Im}\, T_n^{ab} ,\quad
    g_n^{ab}      =   \mathrm{Re}\, T_n^{ab} 
    \label{eq:QG} , \\
    \tilde{T}_n^{ab} &= \Braket{\partial_a n|\left( \hat{H}(\bm{k})-\epsilon_n \right)|\partial_b n} ,\quad
    {T}_n^{ab}        = \Braket{\partial_a n|\hat{Q}_n|\partial_b n} = \Braket{D_a n|D_b n} 
    \label{QGT} .
\end{align}
$\Omega_{nm}^{ab}$ in Eqs.~\eqref{eq:Bijk} and \eqref{eq:Cijk} denotes the band-resolved contribution, with $\Omega_n^{ab}=\sum_{m}^{\neq n}\Omega_{nm}^{ab}$.
These quantities characterize the geometry of the eigenstate space of the Bloch Hamiltonian%
~\cite{provost:cmp1980,berry:book1989,resta:epjb2011,resta:arxiv2017DWOAM,resta:jpcm2018DWSCW,hetenyi:pra2023,kang:natphys2024}.
$m_n^{ab}$ is related to
the orbital magnetic moment of the wave packet
in the semiclassical dynamics $m_n^i$: $m_n^i=e\epsilon_{iab}\, m_n^{ab}/2$,
which describes the self-rotation of the semiclassical wave packet of Bloch electrons around its center%
~\cite{chang:prb1996, sundaram:prb1999, xiao:prl2005, xiao:prl2006thermoele, shi:prl2007, xiao:rmp2010}.
$\Omega_n^{ab}$ and $g_n^{ab}$ are the Berry curvature and the quantum metric, respectively.
As in the expression for the orbital MD%
~\cite{xiao:prl2005, xiao:prl2006thermoele, shi:prl2007, xiao:rmp2010, thonhauser:ijmpb2011, vanderbilt:book}, 
the orbital magnetic moment of the wave packet and the Berry curvature
appear explicitly in the formula because the MO has the same spacetime symmetry as the MD. 
Since the rank of the MO is larger than that of the MD by two, 
quantities with rank two and even parity under spacetime inversion, 
such as the quantum metric and the second momentum derivative, 
appear in the expression to compensate for this difference.
The Berry curvature quadrupole $\partial_j \partial_k \Omega_n^{ab}$ and 
the related quantity $\partial_j \partial_k m_n^{ab}$ appear for the same reason. 
Similar structures also appear in the spin MO, where geometric quantities 
such as the orbital magnetic moment and the Berry curvature are 
replaced by the band-diagonal elements of the spin operator~\cite{oike:prb2025spinMO,TS:npjqmat2026}.
The remaining terms are structurally complicated, however,
these are transformed as
time-reversal-odd rank-3 axial tensors, which correspond to the symmetry of the MO.

\subsection{Relations to response tensors}
\label{subsec:response}
The orbital MO is directly related to a variety of electromagnetic response tensors.
A comprehensive treatment including the spin contribution and lower-rank magnetic multipoles has been provided in the recent study%
~\cite{TS:npjqmat2026}; here, we focus on the orbital MO
and briefly summarize its relations to response functions.

For later convenience, we introduce the derivatives of the MO with respect to $\mu$ as
\begin{align}
    \alpha_{ijk}   &\coloneqq e   \frac{\partial M_{ijk}}{\partial \mu}     ,\quad 
    \alpha'_{ijk}   \coloneqq e^2 \frac{\partial^2 M_{ijk}}{\partial \mu^2} ,\quad 
    \alpha''_{ijk}  \coloneqq e^3 \frac{\partial^3 M_{ijk}}{\partial \mu^3}
    \label{eq:alpha_MO} ,
\end{align}
which naturally appear in linear and nonlinear response functions, as will be shown.
To proceed with the rigorous discussion, we restict ourselves to insulating systems 
at zero temperature ($T=0$).
By utilizing the Maxwell relation in Eq.~\eqref{eq:maxwell},
we can write the polarization charge induced by the magnetic field gradient as
\begin{align}
    e \Delta N &= - \partial_{r_k} P_{k} , \quad
    P_{k} = - \alpha_{ijk} \partial_{r_j} B_{i} ,
    \label{eq:pola_charge}
\end{align}
where $\bm{P}$ denotes the polarization.
We consider centrosymmetric materials, for which a polarization driven solely by a uniform magnetic field is forbidden%
~\cite{gao:prb2018spinMQM,shitade:prb2018orbitalMQM,gao:prb2018orbitalMQM,shitade:prb2019spinMQM}.
The tensor $\alpha_{ijk}$ in Eq.~\eqref{eq:pola_charge} therefore characterizes the QME coupling, i.e., 
a polarization response induced by a magnetic field gradient.
Equation~\eqref{eq:pola_charge} indicates the direct relationship between the MO and the QME tensor.
It should be pointed out, however, that a microscopic formulation of polarization responses 
to spatially nonuniform fields requires a careful treatment 
of spatial inhomogeneities that break lattice periodicity~\cite{xiao:prl2009}.

The MO is also related to magnetization responses induced by an electric field gradient~\cite{shitade:prb2018orbitalMQM}.
The magnetization arising from the nonuniform MO is described as
\begin{align}
    M_{i} &= \partial_{r_j} \partial_{r_k} M_{ijk} ,
    \label{eq:magnetization}
\end{align}
where $\bm{M}$ represents the magnetization.
Considering an electric field $\bm{E} = \partial_{\bm{r}} \mu / e$ 
generated by a nonuniform chemical potential, the magnetization can be written as
\begin{align}
    M_{i} &= \alpha_{ijk} \partial_{r_j} E_{k} + \alpha'_{ijk} E_{j} E_{k}
    \label{eq:mag_response} .
\end{align}
The QME tensor $\alpha_{ijk}$ also characterizes the magnetization response to an electric field gradient.
In contrast, the tensor $\alpha'_{ijk}$ characterizes the second-order magnetoelectric effect.

Using Eq.~\eqref{eq:magnetization}, the magnetization current can then be written as
\begin{align}
    J_{a} 
    &= \epsilon_{a b i} \partial_{r_b} M_{i} \nonumber \\
    &= \sigma_{abjk} \partial_{r_b} \partial_{r_j} E_k
    + \sigma'_{abjk} ( \partial_{r_b} E_j E_k + \partial_{r_k} E_b E_j + \partial_{r_j} E_k E_b )
    + \sigma''_{abjk} E_b E_j E_k .
\end{align}
We find
\begin{align}
    \sigma_{abjk}   &= \epsilon_{a b i} \alpha_{ijk}    \label{eq:sigma_abjk},\quad
    \sigma'_{abjk}   = \epsilon_{a b i} \alpha'_{ijk}   ,\quad
    \sigma''_{abjk}  = \epsilon_{a b i} \alpha''_{ijk}  .
\end{align}
The tensors $\sigma_{abjk}$, $\sigma'_{abjk}$, and $\sigma''_{abjk}$ describe charge current responses to nonuniform or nonlinear electric fields.
In particular, $\sigma_{abjk}$ corresponds to the OAH conductivity~\cite{kozii:prl2021}, while $\sigma''_{abjk}$ describes the 
third-order anomalous Hall conductivity~\cite{xiang:prb2023}. 
A higher-rank analogue of the St\v{r}eda formula%
~\cite{streda:jphysc1982,widom:physletta1982,ceresoli:prb2006,xiao:rmp2010}
can be written explicitly for the OAH conductivity as
\begin{align}
    \sigma_{abjk} 
    &= e \epsilon_{abi} \frac{\partial M_{ijk}}{\partial \mu} 
    = e \epsilon_{abi} \frac{\partial N}{\partial [\partial_{r_j} \partial_{r_k} B_i]} ,
    \label{eq:MO_Streda}
\end{align}
where the Maxwell relation~\eqref{eq:maxwell} has been used to obtain the second equality. 
This equation clearly demonstrates the direct relationship between the orbital MO and the OAH effect.

The above analysis establishes direct thermodynamic relationships 
among the orbital MO, the QME effect, and the OAH effect beyond symmetry considerations.
In contrast, such direct relations do not exist for the nonlinear response
tensors $\alpha'_{ijk}$, $\sigma'_{abjk}$, and $\sigma''_{abjk}$, because the
orbital MO depends linearly on $\mu$ in insulating systems at $T=0$. 
From a thermodynamic viewpoint, the orbital MO is therefore fundamentally
associated with linear responses induced by nonuniform electromagnetic
fields, rather than with nonlinear responses such as the second-order
magnetoelectric effect or the third-order anomalous Hall effect.
It should be emphasized that symmetry considerations alone cannot distinguish
between nonuniform and nonlinear responses. 
We also note that the present OAH effect is qualitatively different 
from the anomalous Hall effect based on the magnetization-space multipoles, 
where the latter explains the in-plane anomalous Hall effect in a unified manner~\cite{liu:prx2025}.

In this work, we discussed the OAH effect only from a thermodynamic perspective. 
The current considered here is a magnetization current, which may be interpreted as an edge current in equillibrium. 
As we mentioned in recent study~\cite{TS:npjqmat2026}, 
it should be distinguished from the transport current detected in conventional transport experiments%
~\cite{cooper:prb1997,xiao:prl2006thermoele,gao:prb2018orbitalMQM}. 
Therefore, a theoretical formulation of an experimentally observable OAH response requires careful consideration.
A related issue has been addressed in previous studies on the orbital magnetic quadrupole~\cite{gao:prb2018orbitalMQM}. 
In the study, the experimentally measurable transport current was derived 
by subtracting the magnetization current from the microscopic current 
carried by semiclassical electron wave packets~\cite{xiao:prl2006thermoele}.
The residual terms associated with this subtraction were shown to provide a microscopic connection 
between the second-order anomalous Hall response in metals and the orbital magnetic quadrupole. 
This observation suggests an important future direction for the present theory. 
In our thermodynamic classification~\cite{TS:npjqmat2026}, 
we could not establish a direct relation between nonlinear responses and magnetic multipoles. 
The example of the orbital magnetic quadrupole indicates, however, that this restriction may be avoided in metals, 
where microscopic calculations can reveal relations between magnetic multipoles and nonlinear responses 
beyond what is accessible from thermodynamics alone.
It is therefore an interesting future problem to examine whether a microscopic relation 
can be established between the orbital MO and the third-order anomalous Hall effect in metals. 
Moreover, since the conventional St\v{r}eda formula can be derived from magnetization currents, 
it is natural to expect that a similar structure may underlie 
the higher-rank St\v{r}eda-type relation in Eq.~\eqref{eq:MO_Streda}.
In other words, just as the quantum Hall effect can be understood from 
both bulk electronic states~\cite{thouless:prl1982TKNN} and edge states~\cite{halperin:prb1982}, 
the higher-rank Hall responses may also be understood from these two complementary perspectives.
A microscopic derivation of the transport current would make it possible to define an experimentally observable OAH conductivity 
and to test whether the higher-rank St\v{r}eda-type relation also holds at the level of transport currents. 
We leave this detailed analysis for future work and will discuss it elsewhere.

\subsection{Model calculations}
\label{subsec:model_calc}
To illustrate the typical properties of the obtained orbital MO, 
we consider two magnetic models with different microscopic origins of time-reversal-symmetry breaking. 
One is a spinless model with orbital magnetic order, 
and the other is a well-known two-sublattice AFM model with momentum-dependent spin splitting.

We first discuss the spinless model. 
To construct a minimal model that hosts a finite orbital MO, 
we consider a two-orbital model on a two-dimensional square lattice, 
inspired by spinless Chern insulator models such as 
the Haldane model~\cite{haldane:prl1988} and the QWZ model~\cite{qi:prb2006QWZ}. 
We take the atomic $d_{x^2-y^2}$ and $d_{xy}$ orbitals localized at the each lattice point as the basis of the Hamiltonian. 
We also impose invariance of the Hamiltonian under the following symmetry operations: 
$\mathcal{P}$, $\mathcal{M}_x$, $\mathcal{M}_y$, $\mathcal{M}_z$, and $C_{4z}\mathcal{T}$.
Here, $\mathcal{P}$ denotes the spatial inversion, 
$\mathcal{M}_i\ (i=x,y,z)$ denotes the mirror operation, and 
$C_{4z}\mathcal{T}$ is an antiunitary operation composed of 
a fourfold rotation $C_{4z}$ about the $z$-axis and the time-reversal operation $\mathcal{T}$.
Under these symmetry constraints, we consider the following Hamiltonian:
\begin{align}
    \hat{H}_{\bm{k}} &= \bm{d}(\bm{k}) \cdot \hat{\bm{\tau}} 
    \label{eq:Hk_for_orbAM} ,\\
    d_x(\bm{k}) &= \sin k_x \sin k_y ( \cos k_x - \cos k_y ) 
    \label{eq:dx} ,\\
    d_y(\bm{k}) &= \sin k_x \sin k_y 
    \label{eq:dy} ,\\
    d_z(\bm{k}) &=  m + 2 - \cos k_x - \cos k_y 
    \label{eq:dz} .
\end{align}
Here, $\hat{\tau}_{i}\ (i=x,y,z)$ are the Pauli matrices in the orbital space.
Since the model is spinless, SOC is absent; thus, it provides an explicit SOC-free platform for examining the orbital MO.

The band structure has an insulating gap for $m>0$ or $m<-4$, 
and the gap size is given by $\Delta = 2(|m+2|-2)$.
The Hamiltonian in Eq.~\eqref{eq:Hk_for_orbAM} preserves the symmetry operations: 
$\mathcal{P}=1$, $\mathcal{M}_x=\hat{\tau}_z$, $\mathcal{M}_y=\hat{\tau}_z$, 
$\mathcal{M}_z=1$, and $C_{4z}\mathcal{T}=-\mathcal{K}$, 
whereas it does not preserve the time-reveresal operation $\mathcal{T}=\mathcal{K}$ 
because of the complex hopping term in Eq.~\eqref{eq:Hk_for_orbAM}.
The presence of the three mirror operations forbids a net orbital magnetization in this model. 
Furthermore, taking into account these mirror operations, the antiunitary symmetry, 
and the two-dimensionality of the system, 
the allowed component of the orbital MO is $M_{zxy}$~\cite{xiao:prl2022hkspaceBCP}.

\begin{figure}[h]
    \centering
    \includegraphics[width=\linewidth]{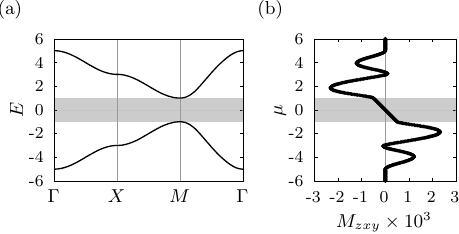}
    \caption{
        Chemical potential dependence of the orbital MO in the spinless two-orbital model. 
        (a) Electronic band structure. 
        (b) $\mu$ dependence of $M_{zxy}$. 
        In (a) and (b), the parameters are set to $m=-5$ and $T=0.01$.
        The shaded areas correspond to the energy gap.
    }
    \label{fig:orbAM_mudep}
\end{figure}

Figures~\ref{fig:orbAM_mudep}(a) and (b) show the electronic band structure and the chemical potential ($\mu$) dependence
of the orbital MO, respectively.
Here we choose $m=-5$ and $T=0.01$, where a small but finite temperature, 
much smaller than the insulating gap, is introduced for numerical stability.
Within the insulating gap, the orbital MO exhibits a linear dependence on $\mu$, 
and the corresponding slope directly yields the QME and OAH response tensor in this model.

\begin{figure}[h]
    \centering
    \includegraphics[width=0.5\linewidth]{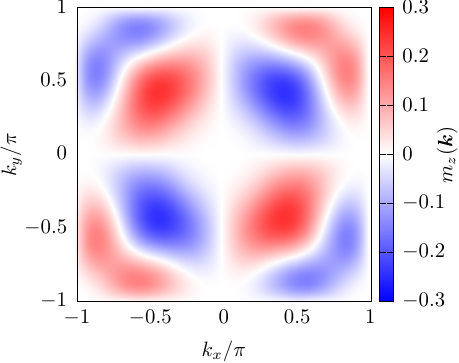}
    \caption{
        Distribution of the orbital magnetic moment $m_z(\bm{k})$ 
        in the Brillouin zone for the spinless two-orbital model. 
        We choose $m=-5$.
    }
    \label{fig:MM_kdist}
\end{figure}

The essential ingredients for realizing a finite orbital MO in this model are 
broken time-reversal symmetry and preserved $C_{4z}\mathcal{T}$ symmetry. 
Importantly, neither spin degrees of freedom nor SOC is required for the finite orbital MO in this model.
This symmetry structure is analogous to that of ordinary altermagnetism, 
where the sublattices in a magnetic unit cell are related by a point-group operation rather than by translation. 
Thus, the present spinless model provides an orbital analogue of spin altermagnetism. 
To gain an intuitive understanding of this orbital altermagnetic character, 
we plot in Fig.~\ref{fig:MM_kdist} the orbital magnetic moment $m_z(\bm{k})$ in the spinless model. 
Here, $m_z(\bm{k}) = \sum_n m_n^z(\bm{k})$, where $m_n^i$ is defined in Sect.~\ref{subsec:MO_formulation}. 
The distribution of $m_z(\bm{k})$ in the Brillouin zone exhibits $d_{xy}$-wave character, 
indicating that the model hosts a finite $M_{zxy}$. 
This behavior also provides an intuitive picture that 
the present spinless model realizes an orbital analogue of spin altermagnetism.
Just as the spin MO characterizes spin altermagnetism and its related responses~\cite{oike:prb2025spinMO,TS:npjqmat2026}, 
the orbital MO characterizes \textsl{orbital altermagnetism} and the corresponding response functions. 
It provides a quantitative characterization of orbital altermagnetism%
~\cite{yu:natcomm2025,sobral:prr2025,pupim:prl2025,chakraborty:arxiv2025orbitalAM,pan:arxiv2025orbitalAM}.
We also note that the mechanism for the finite orbital MO is closely related to 
the Berry-phase mechanism of orbital magnetization~\cite{hoffmann:prb2015,hanke:prb2016}. 
In the present case, for example, complex hoppings that break time-reversal symmetry 
while preserving the relevant antiunitary symmetry can generate finite net Berry curvature quadrupoles 
$\partial_{j}\partial_{k} \Omega_n^{ab}$ 
and the related quantity $\partial_{j}\partial_{k} m_n^{ab}$, 
thereby contributing to the orbital MO even in the absence of SOC.

We then turn to the two-sublattice AFM model. 
We consider a minimal model of the $d_{xy}$-wave AM~\cite{roig:prb2024},
\begin{align}
    \hat{H}_{\bm{k}}
    &= \varepsilon_{0,\bm{k}} + t_{x,\bm{k}} \hat{\tau}_{x} + t_{z,\bm{k}} \hat{\tau}_{z}
    + \lambda_{z,\bm{k}} \hat{\tau}_{y} \hat{\sigma}_z
    + J \hat{\tau}_{z} \hat{\sigma}_z ,
    \label{eq:Hk_for_AM}
\end{align}
where $\hat{\tau}_{i}$ and $\hat{\sigma}_{i}$ are the Pauli matrices for the sublattice and spin degrees of freedom, respectively.
The parameters $t_{x,\bm{k}}$ and $t_{z,\bm{k}}$ denote inter- and intrasublattice hoppings, $\lambda_{z,\bm{k}}$ denotes the relativistic SOC, and $J$ represents the magnetic moment localized at each sublattice. 
The model parameters are chosen to reproduce the nonmagnetic band structure 
of MnF$_2$ reported in Ref.~\cite{roig:prb2024}; 
further details are presented in Sect.~\ref{sec:methods}.
The term $J \hat{\tau}_{z} \hat{\sigma}_z$ in Eq.~\eqref{eq:Hk_for_AM} breaks time-reversal symmetry 
and is essential for realizing finite orbital MO moments.
At the same time, the Hamiltonian preserves the mirror symmetries $\mathcal{M}_x$, $\mathcal{M}_y$, and $\mathcal{M}_z$, 
as well as the antiunitary symmetry $C_{4z} \mathcal{T}$.
These symmetry constraints restrict the allowed components of the orbital MO
to $M_{zxy}$ and $M_{xyz} = M_{yzx}$~\cite{xiao:prl2022hkspaceBCP}.

\begin{figure}[h]
    \centering
    \includegraphics[width=\linewidth]{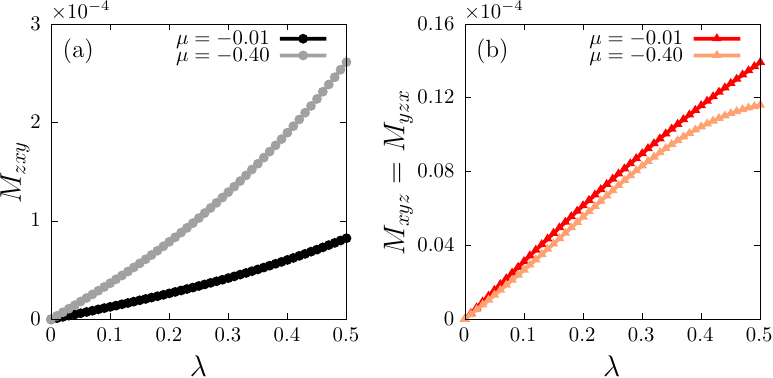}
    \caption{
        SOC dependence of the orbital MO in the spin AM model.
        (a) $M_{zxy}$ and (b) $M_{xyz}=M_{yzx}$ as functions of the SOC strength $\lambda$
        for metallic ($\mu=-0.4$) and insulating ($\mu=-0.01$) regimes.
        The parameters are set to $J=0.5$ and $T=0.01$.
    }
    \label{fig:lamdep}
\end{figure}

Figure~\ref{fig:lamdep} shows the finite components of $M_{ijk}$ in this model as a functions of 
the relativistic SOC strength $\lambda$.
In the numerical simulations, we set $J=0.5$ and $T=0.01$.
Panels (a) and (b) display the results for $M_{zxy}$ and $M_{xyz}=M_{yzx}$, respectively,
for two different values of $\mu$.
For $\mu=-0.4$, the system is metallic, whereas for $\mu=-0.01$ it is insulating.
In both cases, the orbital MOs vanish in the absence of the SOC and increase monotonically with increasing $\lambda$.
These behaviors are in sharp contrast to those of the spin MOs reported previously~\cite{oike:prb2025spinMO, TS:npjqmat2026}, 
for which certain components remain finite even without SOC and show only weak SOC dependence. 
They also differ from those of the spinless two-orbital model discussed above, 
where a finite orbital MO can be obtained without SOC. 
Thus, in the present spin AM model, relativistic SOC is essential for generating the orbital MO.
In this sense, the orbital (spin) MO captures the relativistic (nonrelativistic) properties of the spin AM model.
Equivalently, the relativistic SOC plays a role in converting spin altermagnetism into orbital altermagnetism 
by coupling the altermagnetic spin order to orbital degrees of freedom.

\begin{figure}[h]
    \centering
    \includegraphics[width=\linewidth]{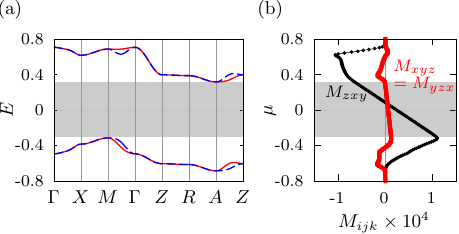}
    \caption{
        Chemical potential dependence of the orbital MO in the spin AM model.
        (a) Electronic band structure.  
        The red solid (blue dashed) lines represent the bands with up(down)-spin polarization.
        (b) $\mu$ dependence of $M_{zxy}$ and $M_{xyz}=M_{yzx}$.
        In (a) and (b), the parameters are set to $\lambda=0.2$, $J=0.5$, and $T=0.01$.
        The shaded areas correspond to the energy gap.
    }
    \label{fig:mudep}
\end{figure}

We further examine the chemical potential dependence of $M_{ijk}$.
Here we fix $\lambda=0.2$, $J=0.5$, and $T=0.01$, where a small but finite temperature, much smaller than the insulating gap, 
is introduced for numerical stability.
Figures~\ref{fig:mudep}(a) and (b) show the electronic band structure and the $\mu$ dependence of the finite components of the orbital MO, respectively. 
Within the insulating gap, the orbital MO exhibits a linear dependence on $\mu$, and the corresponding slope directly yields the QME and OAH response tensor in this model.
We again note that the relativistic SOC is necessary for obtaining the finite orbital responses in this spin AM model.

The OAH effect represents a linear response characteristic of systems 
with ferroically ordered orbital MO moments, as discussed in Sect.~\ref{subsec:response}.
In contrast, the anomalous Hall effect induced by a uniform electric field, often regarded as a hallmark of AMs%
~\cite{solovyev:prb1997,smejkal:sciadv2020,shao:prapp2021,samanta:jappphys2020,sivadas:prl2016,
naka:prb2020,naka:prb2021AHEperovskite,attias:prb2024,chen:prb2022,smejkal:natrevmat2022}, 
is not generally allowed and can be prohibited by symmetry constraints.
In the two models considered here, the conventional anomalous Hall effect is forbidden by 
mirror symmetries~\cite{roig:prb2024}, whereas the OAH effect is symmetry-allowed and finite, 
although a microscopic extraction of the transport current is still required for direct experimental confirmation, 
as discussed in Sect.~\ref{subsec:response}. 
This contrast highlights the OAH effect as a characteristic linear response of AMs. 
We also note that the formula obtained here is applicable to 
a broad class of periodic systems beyond the minimal models considered above.

\section{Conclusions}
\label{sec:conclusions}
In this paper, we have formulated the gauge-invariant expression for the orbital MO 
in crystalline solids based on a thermodynamic approach.
We have clarified the direct relationships between the orbital MO and physical responses 
induced by spatially nonuniform electromagnetic fields, such as the QME and OAH effects.
Through model calculations, we have demonstrated that the orbital MO can characterize orbital altermagnetism, 
which can arise from both orbital magnetic order associated with complex hopping 
and relativistic SOC in conventional spin altermagnetism. 
We also found that the OAH response can remain finite 
even when the conventional anomalous Hall effect is symmetry-forbidden.
These results identify the orbital MO as a key quantity governing nonuniform linear responses in magnetic materials. 
Our formulation is applicable to general periodic systems and 
provides a basis for exploring higher-rank multipole physics and its associated physical phenomena.

\section{Methods}
\label{sec:methods}
The coeffecients used in Eq.~\eqref{eq:Hk_for_AM} are described as~\cite{roig:prb2024}
\begin{align}
    \varepsilon_{0,\bm{k}} &=
    t_1 ( \cos k_x + \cos k_y ) + t_2 \cos k_z + t_3 \cos k_x \cos k_y \nonumber \\
    &\hspace{1em}+t_4 ( \cos k_x + \cos k_y ) \cos k_z + t_5 \cos k_x \cos k_y \cos k_z , \\
    t_{x,\bm{k}}
    &= t_8 \cos \frac{k_x}{2} \cos \frac{k_y}{2} \cos \frac{k_z}{2} , \\
    t_{z,\bm{k}}
    &= t_6 \sin k_x \sin k_y + t_7 \sin k_x \sin k_y \cos k_z , \\
    \lambda_{z,\bm{k}}
    &= \lambda \cos \frac{k_z}{2} \cos \frac{k_x}{2} \cos \frac{k_y}{2}
    (\cos k_x - \cos k_y) .
\end{align}
In the numerical simulations, we adopt the following parameters:
$t_1=0$, $t_2=0.13$, $t_3=0$, $t_4=-0.02$, $t_5=0.015$,
$t_6=0$, $t_7=0.03$, and $t_8=0.33$.
The remaining parameters are specified in Sect.~\ref{sec:results}.

\section*{Abbreviations}
\begin{itemize}
    \item FM \ Ferromagnet 
    \item AFM \ Antiferromagnet
    \item MD \ Magnetic dipole
    \item FM \ Ferromagnet
    \item AM \ Altermagnet
    \item MO \ Magnetic octupole
    \item QME \ Quadrupolar magnetoelectric
    \item OAH \ Octupolar anomalous Hall
    \item SOC \ Spin-orbit coupling
\end{itemize}

\section*{Declarations}
\subsection*{Availability of data and materials}
The datasets used and/or analysed during the current study are available from the corresponding author on reasonable request.
\subsection*{Competing interests}
The authors declare that they have no competing interests.
\subsection*{Funding}
This research was supported by JSPS KAKENHI Grants Numbers JP21H01037, JP22H00101, JP22H01183, JP23H04869, JP23K03288, JP23K20827, and by JST CREST (JPMJCR23O4) and JST FOREST (JPMJFR2366).
\subsection*{Authors' contributions}
T.S. and S.H. conceived the project. 
T.S. performed the analytical and numerical calculations.
Both authors contributed to writing the paper.
\subsection*{Acknowledgements}
The authors thank Kazumasa Hattori and Hiroaki Kusunose for useful comments on this study.
T.S. is also grateful to Di Xiao for pointing out the possible relationship 
between the orbital MO and the third-order anomalous Hall effect.




\begin{thebibliography}{10}
\providecommand{\url}[1]{{#1}}
\providecommand{\urlprefix}{URL }
\providecommand{\doi}[1]{\url{https://doi.org/#1}}


\bibitem{spaldin:jpcm2008}
N.A. Spaldin, M.~Fiebig, M.~Mostovoy, {The toroidal moment in condensed-matter
  physics and its relation to the magnetoelectric effect}.
\newblock Journal of Physics: Condensed Matter \textbf{20}(43), 434203 (2008).
\newblock \doi{10.1088/0953-8984/20/43/434203}.
\newblock \urlprefix\url{https://dx.doi.org/10.1088/0953-8984/20/43/434203}

\bibitem{kusunose:jpsj2008}
H.~Kusunose, {Description of Multipole in f-Electron Systems}.
\newblock Journal of the Physical Society of Japan \textbf{77}(6), 064710
  (2008).
\newblock \doi{10.1143/JPSJ.77.064710}.
\newblock \urlprefix\url{https://doi.org/10.1143/JPSJ.77.064710}.
\newblock
  {\href{https://arxiv.org/abs/https://doi.org/10.1143/JPSJ.77.064710}{{https://doi.org/10.1143/JPSJ.77.064710}}}

\bibitem{kuramoto:ptp2008review}
Y.~Kuramoto, {Electronic Higher Multipoles in Solids}.
\newblock Progress of Theoretical Physics Supplement \textbf{176}, 77--96
  (2008).
\newblock \doi{10.1143/PTPS.176.77}.
\newblock \urlprefix\url{https://doi.org/10.1143/PTPS.176.77}.
\newblock
  {\href{https://arxiv.org/abs/https://academic.oup.com/ptps/article-pdf/doi/10.1143/PTPS.176.77/5324654/176-77.pdf}{{https://academic.oup.com/ptps/article-pdf/doi/10.1143/PTPS.176.77/5324654/176-77.pdf}}}

\bibitem{kuramoto:jpsj2009review}
Y.~Kuramoto, H.~Kusunose, A.~Kiss, {Multipole Orders and Fluctuations in
  Strongly Correlated Electron Systems}.
\newblock Journal of the Physical Society of Japan \textbf{78}(7), 072001
  (2009).
\newblock \doi{10.1143/JPSJ.78.072001}.
\newblock \urlprefix\url{https://doi.org/10.1143/JPSJ.78.072001}.
\newblock
  {\href{https://arxiv.org/abs/https://doi.org/10.1143/JPSJ.78.072001}{{https://doi.org/10.1143/JPSJ.78.072001}}}

\bibitem{santini:rmp2009}
P.~Santini, S.~Carretta, G.~Amoretti, R.~Caciuffo, N.~Magnani, G.H. Lander,
  {Multipolar interactions in $f$-electron systems: The paradigm of actinide
  dioxides}.
\newblock Rev. Mod. Phys. \textbf{81}, 807--863 (2009).
\newblock \doi{10.1103/RevModPhys.81.807}.
\newblock \urlprefix\url{https://link.aps.org/doi/10.1103/RevModPhys.81.807}

\bibitem{watanabe:prb2017}
H.~Watanabe, Y.~Yanase, {Magnetic hexadecapole order and magnetopiezoelectric
  metal state in
  ${\mathbf{Ba}}_{1\ensuremath{-}x}{\mathbf{K}}_{x}{\mathbf{Mn}}_{2}{\mathbf{As}}_{2}$}.
\newblock Phys. Rev. B \textbf{96}, 064432 (2017).
\newblock \doi{10.1103/PhysRevB.96.064432}.
\newblock \urlprefix\url{https://link.aps.org/doi/10.1103/PhysRevB.96.064432}

\bibitem{suzuki:jpsj2018review}
M.T. Suzuki, H.~Ikeda, P.M. Oppeneer, {First-principles Theory of Magnetic
  Multipoles in Condensed Matter Systems}.
\newblock Journal of the Physical Society of Japan \textbf{87}(4), 041008
  (2018).
\newblock \doi{10.7566/JPSJ.87.041008}.
\newblock \urlprefix\url{https://doi.org/10.7566/JPSJ.87.041008}.
\newblock
  {\href{https://arxiv.org/abs/https://doi.org/10.7566/JPSJ.87.041008}{{https://doi.org/10.7566/JPSJ.87.041008}}}

\bibitem{hayami:prb2018classification}
S.~Hayami, M.~Yatsushiro, Y.~Yanagi, H.~Kusunose, {Classification of
  atomic-scale multipoles under crystallographic point groups and application
  to linear response tensors}.
\newblock Phys. Rev. B \textbf{98}, 165110 (2018).
\newblock \doi{10.1103/PhysRevB.98.165110}.
\newblock \urlprefix\url{https://link.aps.org/doi/10.1103/PhysRevB.98.165110}

\bibitem{watanabe:prb2018classification}
H.~Watanabe, Y.~Yanase, {Group-theoretical classification of multipole order:
  Emergent responses and candidate materials}.
\newblock Phys. Rev. B \textbf{98}, 245129 (2018).
\newblock \doi{10.1103/PhysRevB.98.245129}.
\newblock \urlprefix\url{https://link.aps.org/doi/10.1103/PhysRevB.98.245129}

\bibitem{yatsushiro:prb2021classification}
M.~Yatsushiro, H.~Kusunose, S.~Hayami, {Multipole classification in 122
  magnetic point groups for unified understanding of multiferroic responses and
  transport phenomena}.
\newblock Phys. Rev. B \textbf{104}, 054412 (2021).
\newblock \doi{10.1103/PhysRevB.104.054412}.
\newblock \urlprefix\url{https://link.aps.org/doi/10.1103/PhysRevB.104.054412}

\bibitem{hayami:jpscp2020}
S.~Hayami, Y.~Yanagi, M.~Naka, H.~Seo, Y.~Motome, H.~Kusunose, {Multipole
  Description of Emergent Spin-Orbit Interaction in Organic Antiferromagnet
  \(\kappa \)-(BEDT-TTF)$_2$Cu[N(CN)$_2$]Cl}.
\newblock JPS Conf. Proc. \textbf{30}, 011149 (2020).
\newblock \doi{10.7566/JPSCP.30.011149}.
\newblock
  \urlprefix\url{https://journals.jps.jp/doi/abs/10.7566/JPSCP.30.011149}.
\newblock
  {\href{https://arxiv.org/abs/https://journals.jps.jp/doi/pdf/10.7566/JPSCP.30.011149}{{https://journals.jps.jp/doi/pdf/10.7566/JPSCP.30.011149}}}

\bibitem{hayami:jpsj2019spinsplit}
S.~Hayami, Y.~Yanagi, H.~Kusunose, {Momentum-Dependent Spin Splitting by
  Collinear Antiferromagnetic Ordering}.
\newblock Journal of the Physical Society of Japan \textbf{88}(12), 123702
  (2019).
\newblock \doi{10.7566/JPSJ.88.123702}.
\newblock \urlprefix\url{https://doi.org/10.7566/JPSJ.88.123702}.
\newblock
  {\href{https://arxiv.org/abs/https://doi.org/10.7566/JPSJ.88.123702}{{https://doi.org/10.7566/JPSJ.88.123702}}}

\bibitem{hayami:prb2020antisymspinsplit}
S.~Hayami, Y.~Yanagi, H.~Kusunose, {Spontaneous antisymmetric spin splitting in
  noncollinear antiferromagnets without spin-orbit coupling}.
\newblock Phys. Rev. B \textbf{101}, 220403 (2020).
\newblock \doi{10.1103/PhysRevB.101.220403}.
\newblock \urlprefix\url{https://link.aps.org/doi/10.1103/PhysRevB.101.220403}

\bibitem{hayami:prb2020spinsplit}
S.~Hayami, Y.~Yanagi, H.~Kusunose, {Bottom-up design of spin-split and reshaped
  electronic band structures in antiferromagnets without spin-orbit coupling:
  Procedure on the basis of augmented multipoles}.
\newblock Phys. Rev. B \textbf{102}, 144441 (2020).
\newblock \doi{10.1103/PhysRevB.102.144441}.
\newblock \urlprefix\url{https://link.aps.org/doi/10.1103/PhysRevB.102.144441}

\bibitem{hayami:jpsj2024review}
S.~Hayami, H.~Kusunose, {Unified Description of Electronic Orderings and Cross
  Correlations by Complete Multipole Representation}.
\newblock Journal of the Physical Society of Japan \textbf{93}(7), 072001
  (2024).
\newblock \doi{10.7566/JPSJ.93.072001}.
\newblock \urlprefix\url{https://doi.org/10.7566/JPSJ.93.072001}.
\newblock
  {\href{https://arxiv.org/abs/https://doi.org/10.7566/JPSJ.93.072001}{{https://doi.org/10.7566/JPSJ.93.072001}}}

\bibitem{watanabe:jpcm2024review}
H.~Watanabe, Y.~Yanase, {Magnetic parity violation and
  parity-time-reversal-symmetric magnets}.
\newblock Journal of Physics: Condensed Matter \textbf{36}(37), 373001 (2024).
\newblock \doi{10.1088/1361-648X/ad52dd}.
\newblock \urlprefix\url{https://doi.org/10.1088/1361-648X/ad52dd}

\bibitem{noda:pccp2016}
Y.~Noda, K.~Ohno, S.~Nakamura, {Momentum-dependent band spin splitting in
  semiconducting MnO2: a density functional calculation}.
\newblock Phys. Chem. Chem. Phys. \textbf{18}, 13294--13303 (2016).
\newblock \doi{10.1039/C5CP07806G}.
\newblock \urlprefix\url{http://dx.doi.org/10.1039/C5CP07806G}

\bibitem{okugawa:jpcm2018}
T.~Okugawa, K.~Ohno, Y.~Noda, S.~Nakamura, {Weakly spin-dependent band
  structures of antiferromagnetic perovskite LaMO3 (M = Cr, Mn, Fe)}.
\newblock Journal of Physics: Condensed Matter \textbf{30}(7), 075502 (2018).
\newblock \doi{10.1088/1361-648X/aa9e70}.
\newblock \urlprefix\url{https://dx.doi.org/10.1088/1361-648X/aa9e70}

\bibitem{smejkal:sciadv2020}
L.~Šmejkal, R.~González-Hernández, T.~Jungwirth, J.~Sinova, {Crystal
  time-reversal symmetry breaking and spontaneous Hall effect in collinear
  antiferromagnets}.
\newblock Science Advances \textbf{6}(23), eaaz8809 (2020).
\newblock \doi{10.1126/sciadv.aaz8809}.
\newblock
  \urlprefix\url{https://www.science.org/doi/abs/10.1126/sciadv.aaz8809}.
\newblock
  {\href{https://arxiv.org/abs/https://www.science.org/doi/pdf/10.1126/sciadv.aaz8809}{{https://www.science.org/doi/pdf/10.1126/sciadv.aaz8809}}}

\bibitem{naka:natcomm2019}
M.~Naka, S.~Hayami, H.~Kusunose, Y.~Yanagi, Y.~Motome, H.~Seo, {Spin current
  generation in organic antiferromagnets}.
\newblock Nature Communications \textbf{10}, 4305 (2019).
\newblock \doi{10.1038/s41467-019-12229-y}.
\newblock \urlprefix\url{https://doi.org/10.1038/s41467-019-12229-y}

\bibitem{ahn:prb2019}
K.H. Ahn, A.~Hariki, K.W. Lee, J.~Kune\ifmmode~\check{s}\else \v{s}\fi{},
  {Antiferromagnetism in ${\mathrm{RuO}}_{2}$ as $d$-wave Pomeranchuk
  instability}.
\newblock Phys. Rev. B \textbf{99}, 184432 (2019).
\newblock \doi{10.1103/PhysRevB.99.184432}.
\newblock \urlprefix\url{https://link.aps.org/doi/10.1103/PhysRevB.99.184432}

\bibitem{yuan:prb2020}
L.D. Yuan, Z.~Wang, J.W. Luo, E.I. Rashba, A.~Zunger, {Giant momentum-dependent
  spin splitting in centrosymmetric low-$Z$ antiferromagnets}.
\newblock Phys. Rev. B \textbf{102}, 014422 (2020).
\newblock \doi{10.1103/PhysRevB.102.014422}.
\newblock \urlprefix\url{https://link.aps.org/doi/10.1103/PhysRevB.102.014422}

\bibitem{naka:prb2021perovskite}
M.~Naka, Y.~Motome, H.~Seo, {Perovskite as a spin current generator}.
\newblock Phys. Rev. B \textbf{103}, 125114 (2021).
\newblock \doi{10.1103/PhysRevB.103.125114}.
\newblock \urlprefix\url{https://link.aps.org/doi/10.1103/PhysRevB.103.125114}

\bibitem{yuan:prm2021}
L.D. Yuan, Z.~Wang, J.W. Luo, A.~Zunger, {Prediction of low-Z collinear and
  noncollinear antiferromagnetic compounds having momentum-dependent spin
  splitting even without spin-orbit coupling}.
\newblock Phys. Rev. Mater. \textbf{5}, 014409 (2021).
\newblock \doi{10.1103/PhysRevMaterials.5.014409}.
\newblock
  \urlprefix\url{https://link.aps.org/doi/10.1103/PhysRevMaterials.5.014409}

\bibitem{yuan:prb2021}
L.D. Yuan, Z.~Wang, J.W. Luo, A.~Zunger, {Strong influence of nonmagnetic
  ligands on the momentum-dependent spin splitting in antiferromagnets}.
\newblock Phys. Rev. B \textbf{103}, 224410 (2021).
\newblock \doi{10.1103/PhysRevB.103.224410}.
\newblock \urlprefix\url{https://link.aps.org/doi/10.1103/PhysRevB.103.224410}

\bibitem{hernandez:prl2021}
R.~Gonz\'alez-Hern\'andez, L.~\ifmmode~\check{S}\else \v{S}\fi{}mejkal,
  K.~V\'yborn\'y, Y.~Yahagi, J.~Sinova, T.c.v. Jungwirth,
  J.~\ifmmode~\check{Z}\else \v{Z}\fi{}elezn\'y, {Efficient Electrical Spin
  Splitter Based on Nonrelativistic Collinear Antiferromagnetism}.
\newblock Phys. Rev. Lett. \textbf{126}, 127701 (2021).
\newblock \doi{10.1103/PhysRevLett.126.127701}.
\newblock
  \urlprefix\url{https://link.aps.org/doi/10.1103/PhysRevLett.126.127701}

\bibitem{smejkal:prx2022magnetoresistance}
L.~\ifmmode~\check{S}\else \v{S}\fi{}mejkal, A.B. Hellenes,
  R.~Gonz\'alez-Hern\'andez, J.~Sinova, T.~Jungwirth, {Giant and Tunneling
  Magnetoresistance in Unconventional Collinear Antiferromagnets with
  Nonrelativistic Spin-Momentum Coupling}.
\newblock Phys. Rev. X \textbf{12}, 011028 (2022).
\newblock \doi{10.1103/PhysRevX.12.011028}.
\newblock \urlprefix\url{https://link.aps.org/doi/10.1103/PhysRevX.12.011028}

\bibitem{mazin:pnas2021}
I.I. Mazin, K.~Koepernik, M.D. Johannes, R.~Gonz{\'a}lez-Hern{\'a}ndez,
  L.~{\v{S}}mejkal, {Prediction of unconventional magnetism in doped FeSb$_2$}.
\newblock Proceedings of the National Academy of Sciences \textbf{118}(42),
  e2108924118 (2021)

\bibitem{smejkal:prx2022spingroup1}
L.~\ifmmode~\check{S}\else \v{S}\fi{}mejkal, J.~Sinova, T.~Jungwirth, {Beyond
  Conventional Ferromagnetism and Antiferromagnetism: A Phase with
  Nonrelativistic Spin and Crystal Rotation Symmetry}.
\newblock Phys. Rev. X \textbf{12}, 031042 (2022).
\newblock \doi{10.1103/PhysRevX.12.031042}.
\newblock \urlprefix\url{https://link.aps.org/doi/10.1103/PhysRevX.12.031042}

\bibitem{smejkal:prx2022spingroup2}
L.~\ifmmode~\check{S}\else \v{S}\fi{}mejkal, J.~Sinova, T.~Jungwirth, {Emerging
  Research Landscape of Altermagnetism}.
\newblock Phys. Rev. X \textbf{12}, 040501 (2022).
\newblock \doi{10.1103/PhysRevX.12.040501}.
\newblock \urlprefix\url{https://link.aps.org/doi/10.1103/PhysRevX.12.040501}

\bibitem{cheong:npj2025AMclassification}
S.W. Cheong, F.T. Huang, {Altermagnetism classification}.
\newblock npj Quantum Materials \textbf{10}(1), 38 (2025)

\bibitem{guo:advmat2025review}
Z.~Guo, X.~Wang, W.~Wang, G.~Zhang, X.~Zhou, Z.~Cheng, {Spin-Polarized
  Antiferromagnets for Spintronics}.
\newblock Advanced Materials p. 2505779 (2025)

\bibitem{hu:prx2025}
M.~Hu, X.~Cheng, Z.~Huang, J.~Liu, {Catalog of $C$-Paired Spin-Momentum Locking
  in Antiferromagnetic Systems}.
\newblock Phys. Rev. X \textbf{15}, 021083 (2025).
\newblock \doi{10.1103/PhysRevX.15.021083}.
\newblock \urlprefix\url{https://link.aps.org/doi/10.1103/PhysRevX.15.021083}

\bibitem{bhowal:prx2024}
S.~Bhowal, N.A. Spaldin, {Ferroically Ordered Magnetic Octupoles in $d$-Wave
  Altermagnets}.
\newblock Phys. Rev. X \textbf{14}, 011019 (2024).
\newblock \doi{10.1103/PhysRevX.14.011019}.
\newblock \urlprefix\url{https://link.aps.org/doi/10.1103/PhysRevX.14.011019}

\bibitem{jackson}
J.D. Jackson, \emph{{Classical Electrodynamics}}, 3rd edn. (Wiley, New York,
  1999)

\bibitem{king-smith:prb1993}
R.D. King-Smith, D.~Vanderbilt, {Theory of polarization of crystalline solids}.
\newblock Phys. Rev. B \textbf{47}, 1651--1654 (1993).
\newblock \doi{10.1103/PhysRevB.47.1651}.
\newblock \urlprefix\url{https://link.aps.org/doi/10.1103/PhysRevB.47.1651}

\bibitem{vanderbilt:prb1993}
D.~Vanderbilt, R.D. King-Smith, {Electric polarization as a bulk quantity and
  its relation to surface charge}.
\newblock Phys. Rev. B \textbf{48}, 4442--4455 (1993).
\newblock \doi{10.1103/PhysRevB.48.4442}.
\newblock \urlprefix\url{https://link.aps.org/doi/10.1103/PhysRevB.48.4442}

\bibitem{resta:rmp1994}
R.~Resta, {Macroscopic polarization in crystalline dielectrics: the geometric
  phase approach}.
\newblock Rev. Mod. Phys. \textbf{66}, 899--915 (1994).
\newblock \doi{10.1103/RevModPhys.66.899}.
\newblock \urlprefix\url{https://link.aps.org/doi/10.1103/RevModPhys.66.899}

\bibitem{resta:book}
R.~Resta, D.~Vanderbilt, \emph{{Theory of Polarization: A Modern Approach}}
  (Springer Berlin Heidelberg, Berlin, Heidelberg, 2007), pp. 31--68.
\newblock \doi{10.1007/978-3-540-34591-6_2}.
\newblock \urlprefix\url{https://doi.org/10.1007/978-3-540-34591-6_2}

\bibitem{resta:jpcm2010review}
R.~Resta, {Electrical polarization and orbital magnetization: the modern
  theories}.
\newblock Journal of Physics: Condensed Matter \textbf{22}(12), 123201 (2010).
\newblock \doi{10.1088/0953-8984/22/12/123201}.
\newblock \urlprefix\url{https://dx.doi.org/10.1088/0953-8984/22/12/123201}

\bibitem{xiao:rmp2010}
D.~Xiao, M.C. Chang, Q.~Niu, {Berry phase effects on electronic properties}.
\newblock Rev. Mod. Phys. \textbf{82}, 1959--2007 (2010).
\newblock \doi{10.1103/RevModPhys.82.1959}.
\newblock \urlprefix\url{https://link.aps.org/doi/10.1103/RevModPhys.82.1959}

\bibitem{vanderbilt:book}
D.~Vanderbilt.
\newblock {Berry Phases in Electronic Structure Theory: Electric Polarization,
  Orbital Magnetization and Topological Insulators} (Cambridge University
  Press, 2018)

\bibitem{resta:chemphyschem2005}
R.~Resta, D.~Ceresoli, T.~Thonhauser, D.~Vanderbilt, {Orbital Magnetization in
  Extended Systems}.
\newblock ChemPhysChem \textbf{6}(9), 1815--1819 (2005).
\newblock \doi{https://doi.org/10.1002/cphc.200400641}.
\newblock
  \urlprefix\url{https://chemistry-europe.onlinelibrary.wiley.com/doi/abs/10.1002/cphc.200400641}.
\newblock
  {\href{https://arxiv.org/abs/https://chemistry-europe.onlinelibrary.wiley.com/doi/pdf/10.1002/cphc.200400641}{{https://chemistry-europe.onlinelibrary.wiley.com/doi/pdf/10.1002/cphc.200400641}}}

\bibitem{xiao:prl2005}
D.~Xiao, J.~Shi, Q.~Niu, {Berry Phase Correction to Electron Density of States
  in Solids}.
\newblock Phys. Rev. Lett. \textbf{95}, 137204 (2005).
\newblock \doi{10.1103/PhysRevLett.95.137204}.
\newblock
  \urlprefix\url{https://link.aps.org/doi/10.1103/PhysRevLett.95.137204}

\bibitem{xiao:prl2006thermoele}
D.~Xiao, Y.~Yao, Z.~Fang, Q.~Niu, {Berry-Phase Effect in Anomalous
  Thermoelectric Transport}.
\newblock Phys. Rev. Lett. \textbf{97}, 026603 (2006).
\newblock \doi{10.1103/PhysRevLett.97.026603}.
\newblock
  \urlprefix\url{https://link.aps.org/doi/10.1103/PhysRevLett.97.026603}

\bibitem{thonhauser:prl2005}
T.~Thonhauser, D.~Ceresoli, D.~Vanderbilt, R.~Resta, {Orbital Magnetization in
  Periodic Insulators}.
\newblock Phys. Rev. Lett. \textbf{95}, 137205 (2005).
\newblock \doi{10.1103/PhysRevLett.95.137205}.
\newblock
  \urlprefix\url{https://link.aps.org/doi/10.1103/PhysRevLett.95.137205}

\bibitem{ceresoli:prb2006}
D.~Ceresoli, T.~Thonhauser, D.~Vanderbilt, R.~Resta, {Orbital magnetization in
  crystalline solids: Multi-band insulators, Chern insulators, and metals}.
\newblock Phys. Rev. B \textbf{74}, 024408 (2006).
\newblock \doi{10.1103/PhysRevB.74.024408}.
\newblock \urlprefix\url{https://link.aps.org/doi/10.1103/PhysRevB.74.024408}

\bibitem{shi:prl2007}
J.~Shi, G.~Vignale, D.~Xiao, Q.~Niu, {Quantum Theory of Orbital Magnetization
  and Its Generalization to Interacting Systems}.
\newblock Phys. Rev. Lett. \textbf{99}, 197202 (2007).
\newblock \doi{10.1103/PhysRevLett.99.197202}.
\newblock
  \urlprefix\url{https://link.aps.org/doi/10.1103/PhysRevLett.99.197202}

\bibitem{souza:prb2008}
I.~Souza, D.~Vanderbilt, {Dichroic $f$-sum rule and the orbital magnetization
  of crystals}.
\newblock Phys. Rev. B \textbf{77}, 054438 (2008).
\newblock \doi{10.1103/PhysRevB.77.054438}.
\newblock \urlprefix\url{https://link.aps.org/doi/10.1103/PhysRevB.77.054438}

\bibitem{thonhauser:ijmpb2011}
T.~THONHAUSER, {THEORY OF ORBITAL MAGNETIZATION IN SOLIDS}.
\newblock International Journal of Modern Physics B \textbf{25}(11), 1429--1458
  (2011).
\newblock \doi{10.1142/S0217979211058912}.
\newblock \urlprefix\url{https://doi.org/10.1142/S0217979211058912}.
\newblock
  {\href{https://arxiv.org/abs/https://doi.org/10.1142/S0217979211058912}{{https://doi.org/10.1142/S0217979211058912}}}

\bibitem{bianco:prl2013}
R.~Bianco, R.~Resta, Orbital magnetization as a local property.
\newblock Phys. Rev. Lett. \textbf{110}, 087202 (2013).
\newblock \doi{10.1103/PhysRevLett.110.087202}.
\newblock
  \urlprefix\url{https://link.aps.org/doi/10.1103/PhysRevLett.110.087202}

\bibitem{seleznev:prb2023}
D.~Seleznev, D.~Vanderbilt, Towards a theory of surface orbital magnetization.
\newblock Phys. Rev. B \textbf{107}, 115102 (2023).
\newblock \doi{10.1103/PhysRevB.107.115102}.
\newblock \urlprefix\url{https://link.aps.org/doi/10.1103/PhysRevB.107.115102}

\bibitem{gao:prb2018spinMQM}
Y.~Gao, D.~Vanderbilt, D.~Xiao, {Microscopic theory of spin toroidization in
  periodic crystals}.
\newblock Phys. Rev. B \textbf{97}, 134423 (2018).
\newblock \doi{10.1103/PhysRevB.97.134423}.
\newblock \urlprefix\url{https://link.aps.org/doi/10.1103/PhysRevB.97.134423}

\bibitem{shitade:prb2018orbitalMQM}
A.~Shitade, H.~Watanabe, Y.~Yanase, {Theory of orbital magnetic quadrupole
  moment and magnetoelectric susceptibility}.
\newblock Phys. Rev. B \textbf{98}, 020407 (2018).
\newblock \doi{10.1103/PhysRevB.98.020407}.
\newblock \urlprefix\url{https://link.aps.org/doi/10.1103/PhysRevB.98.020407}

\bibitem{gao:prb2018orbitalMQM}
Y.~Gao, D.~Xiao, {Orbital magnetic quadrupole moment and nonlinear anomalous
  thermoelectric transport}.
\newblock Phys. Rev. B \textbf{98}, 060402 (2018).
\newblock \doi{10.1103/PhysRevB.98.060402}.
\newblock \urlprefix\url{https://link.aps.org/doi/10.1103/PhysRevB.98.060402}

\bibitem{shitade:prb2019spinMQM}
A.~Shitade, A.~Daido, Y.~Yanase, {Theory of spin magnetic quadrupole moment and
  temperature-gradient-induced magnetization}.
\newblock Phys. Rev. B \textbf{99}, 024404 (2019).
\newblock \doi{10.1103/PhysRevB.99.024404}.
\newblock \urlprefix\url{https://link.aps.org/doi/10.1103/PhysRevB.99.024404}

\bibitem{daido:prb2020}
A.~Daido, A.~Shitade, Y.~Yanase, {Thermodynamic approach to electric quadrupole
  moments}.
\newblock Phys. Rev. B \textbf{102}, 235149 (2020).
\newblock \doi{10.1103/PhysRevB.102.235149}.
\newblock \urlprefix\url{https://link.aps.org/doi/10.1103/PhysRevB.102.235149}

\bibitem{oike:prb2025spinMO}
J.~\ifmmode~\bar{O}\else \={O}\fi{}ik\'e, R.~Peters, K.~Shinada, {Thermodynamic
  formulation of the spin magnetic octupole moment in bulk crystals}.
\newblock Phys. Rev. B \textbf{112}, 134412 (2025).
\newblock \doi{10.1103/frq1-9xx7}.
\newblock \urlprefix\url{https://link.aps.org/doi/10.1103/frq1-9xx7}

\bibitem{TS:npjqmat2026}
T.~Sato, S.~Hayami, {Quantum theory of magnetic octupole in periodic crystals
  and application to d-wave altermagnets}.
\newblock npj Quantum Materials \textbf{11}, 32 (2026).
\newblock \doi{https://doi.org/10.1038/s41535-026-00865-9}.
\newblock \urlprefix\url{https://www.nature.com/articles/s41535-026-00865-9}

\bibitem{shitade:prb2025}
A.~Shitade, {Intrinsic spin accumulation in the magnetic spin Hall effect}.
\newblock Phys. Rev. B \textbf{112}, 174431 (2025).
\newblock \doi{10.1103/gxpm-2gkq}.
\newblock \urlprefix\url{https://link.aps.org/doi/10.1103/gxpm-2gkq}

\bibitem{kozii:prl2021}
V.~Kozii, A.~Avdoshkin, S.~Zhong, J.E. Moore, {Intrinsic Anomalous Hall
  Conductivity in a Nonuniform Electric Field}.
\newblock Phys. Rev. Lett. \textbf{126}, 156602 (2021).
\newblock \doi{10.1103/PhysRevLett.126.156602}.
\newblock
  \urlprefix\url{https://link.aps.org/doi/10.1103/PhysRevLett.126.156602}

\bibitem{streda:jphysc1982}
P.~Streda, {Theory of quantised Hall conductivity in two dimensions}.
\newblock Journal of Physics C: Solid State Physics \textbf{15}(22), L717
  (1982).
\newblock \doi{10.1088/0022-3719/15/22/005}.
\newblock \urlprefix\url{https://dx.doi.org/10.1088/0022-3719/15/22/005}

\bibitem{widom:physletta1982}
A.~Widom, {Thermodynamic derivation of the Hall effect current}.
\newblock Physics Letters A \textbf{90}(9), 474 (1982).
\newblock \doi{https://doi.org/10.1016/0375-9601(82)90401-7}.
\newblock
  \urlprefix\url{https://www.sciencedirect.com/science/article/pii/0375960182904017}

\bibitem{yu:natcomm2025}
Y.~Yu, H.G. Suh, M.~Roig, D.F. Agterberg, {Altermagnetism from coincident Van
  Hove singularities: application to $\kappa$-Cl}.
\newblock Nature Communications \textbf{16}(1), 2950 (2025)

\bibitem{sobral:prr2025}
J.a.A. Sobral, S.~Mandal, M.S. Scheurer, {Fractionalized altermagnets: From
  neighboring and altermagnetic spin liquids to spin-symmetric band splitting}.
\newblock Phys. Rev. Res. \textbf{7}, 023152 (2025).
\newblock \doi{10.1103/PhysRevResearch.7.023152}.
\newblock
  \urlprefix\url{https://link.aps.org/doi/10.1103/PhysRevResearch.7.023152}

\bibitem{pupim:prl2025}
L.V. Pupim, M.S. Scheurer, {Adatom Engineering Magnetic Order in
  Superconductors: Applications to Altermagnetic Superconductivity}.
\newblock Phys. Rev. Lett. \textbf{134}, 146001 (2025).
\newblock \doi{10.1103/PhysRevLett.134.146001}.
\newblock
  \urlprefix\url{https://link.aps.org/doi/10.1103/PhysRevLett.134.146001}

\bibitem{chakraborty:arxiv2025orbitalAM}
A.R. Chakraborty, F.~Yang, T.~Birol, R.M. Fernandes.
\newblock {Orbital altermagnetism on the kagome lattice and possible
  application to $A$V$_3$Sb$_5$} (2025).
\newblock \urlprefix\url{https://arxiv.org/abs/2509.26596}

\bibitem{pan:arxiv2025orbitalAM}
M.~Pan, F.~Liu, H.~Huang.
\newblock Orbital altermagnetism (2026).
\newblock \doi{https://doi.org/10.1103/l8fc-dp36}.
\newblock \urlprefix\url{https://arxiv.org/abs/2510.00509}

\bibitem{shitade:prb2019magnonGME}
A.~Shitade, Y.~Yanase, {Magnon gravitomagnetoelectric effect in
  noncentrosymmetric antiferromagnetic insulators}.
\newblock Phys. Rev. B \textbf{100}, 224416 (2019).
\newblock \doi{10.1103/PhysRevB.100.224416}.
\newblock \urlprefix\url{https://link.aps.org/doi/10.1103/PhysRevB.100.224416}

\bibitem{provost:cmp1980}
J.P. Provost, G.~Vallee, {Riemannian structure on manifolds of quantum states}.
\newblock Communications in Mathematical Physics \textbf{76}(3), 289 -- 301
  (1980)

\bibitem{berry:book1989}
M.V. Berry, \emph{{The Quantum Phase, Five Years After}}.
\newblock Advanced series in mathematical physics (World Scientific Publishing
  Company, 1989), pp. 7--28

\bibitem{resta:epjb2011}
R.~Resta, {The insulating state of matter: a geometrical theory}.
\newblock The European Physical Journal B \textbf{79}(2), 121--137 (2011).
\newblock \doi{10.1140/epjb/e2010-10874-4}.
\newblock \urlprefix\url{http://dx.doi.org/10.1140/epjb/e2010-10874-4}

\bibitem{resta:arxiv2017DWOAM}
R.~Resta.
\newblock {Geometrical meaning of the Drude weight and its relationship to
  orbital magnetization} (2017).
\newblock \urlprefix\url{https://arxiv.org/abs/1703.00712}

\bibitem{resta:jpcm2018DWSCW}
R.~Resta, {Drude weight and superconducting weight}.
\newblock Journal of Physics: Condensed Matter \textbf{30}(41), 414001 (2018).
\newblock \doi{10.1088/1361-648X/aade19}.
\newblock \urlprefix\url{https://dx.doi.org/10.1088/1361-648X/aade19}

\bibitem{hetenyi:pra2023}
B.~Het\'enyi, P.~L\'evay, {Fluctuations, uncertainty relations, and the
  geometry of quantum state manifolds}.
\newblock Phys. Rev. A \textbf{108}, 032218 (2023).
\newblock \doi{10.1103/PhysRevA.108.032218}.
\newblock \urlprefix\url{https://link.aps.org/doi/10.1103/PhysRevA.108.032218}

\bibitem{kang:natphys2024}
M.~Kang, S.~Kim, Y.~Qian, P.M. Neves, L.~Ye, J.~Jung, D.~Puntel, F.~Mazzola,
  S.~Fang, C.~Jozwiak, A.~Bostwick, E.~Rotenberg, J.~Fuji, I.~Vobornik, J.H.
  Park, J.G. Checkelsky, B.J. Yang, R.~Comin, {Measurements of the quantum
  geometric tensor in solids}.
\newblock Nature Physics \textbf{21}(1), 110--117 (2024).
\newblock \doi{10.1038/s41567-024-02678-8}.
\newblock \urlprefix\url{http://dx.doi.org/10.1038/s41567-024-02678-8}

\bibitem{chang:prb1996}
M.C. Chang, Q.~Niu, {Berry phase, hyperorbits, and the Hofstadter spectrum:
  Semiclassical dynamics in magnetic Bloch bands}.
\newblock Phys. Rev. B \textbf{53}, 7010--7023 (1996).
\newblock \doi{10.1103/PhysRevB.53.7010}.
\newblock \urlprefix\url{https://link.aps.org/doi/10.1103/PhysRevB.53.7010}

\bibitem{sundaram:prb1999}
G.~Sundaram, Q.~Niu, {Wave-packet dynamics in slowly perturbed crystals:
  Gradient corrections and Berry-phase effects}.
\newblock Phys. Rev. B \textbf{59}, 14915--14925 (1999).
\newblock \doi{10.1103/PhysRevB.59.14915}.
\newblock \urlprefix\url{https://link.aps.org/doi/10.1103/PhysRevB.59.14915}

\bibitem{xiao:prl2009}
D.~Xiao, J.~Shi, D.P. Clougherty, Q.~Niu, {Polarization and Adiabatic Pumping
  in Inhomogeneous Crystals}.
\newblock Phys. Rev. Lett. \textbf{102}, 087602 (2009).
\newblock \doi{10.1103/PhysRevLett.102.087602}.
\newblock
  \urlprefix\url{https://link.aps.org/doi/10.1103/PhysRevLett.102.087602}

\bibitem{xiang:prb2023}
L.~Xiang, C.~Zhang, L.~Wang, J.~Wang, {Third-order intrinsic anomalous Hall
  effect with generalized semiclassical theory}.
\newblock Phys. Rev. B \textbf{107}, 075411 (2023).
\newblock \doi{10.1103/PhysRevB.107.075411}.
\newblock \urlprefix\url{https://link.aps.org/doi/10.1103/PhysRevB.107.075411}

\bibitem{liu:prx2025}
Z.~Liu, M.~Wei, W.~Peng, D.~Hou, Y.~Gao, Q.~Niu, {Multipolar Anisotropy in
  Anomalous Hall Effect from Spin-Group Symmetry Breaking}.
\newblock Phys. Rev. X \textbf{15}, 031006 (2025).
\newblock \doi{10.1103/PhysRevX.15.031006}.
\newblock \urlprefix\url{https://link.aps.org/doi/10.1103/PhysRevX.15.031006}

\bibitem{cooper:prb1997}
N.R. Cooper, B.I. Halperin, I.M. Ruzin, {Thermoelectric response of an
  interacting two-dimensional electron gas in a quantizing magnetic field}.
\newblock Phys. Rev. B \textbf{55}, 2344--2359 (1997).
\newblock \doi{10.1103/PhysRevB.55.2344}.
\newblock \urlprefix\url{https://link.aps.org/doi/10.1103/PhysRevB.55.2344}

\bibitem{thouless:prl1982TKNN}
D.J. Thouless, M.~Kohmoto, M.P. Nightingale, M.~den Nijs, {Quantized Hall
  Conductance in a Two-Dimensional Periodic Potential}.
\newblock Phys. Rev. Lett. \textbf{49}, 405--408 (1982).
\newblock \doi{10.1103/PhysRevLett.49.405}.
\newblock \urlprefix\url{https://link.aps.org/doi/10.1103/PhysRevLett.49.405}

\bibitem{halperin:prb1982}
B.I. Halperin, {Quantized Hall conductance, current-carrying edge states, and
  the existence of extended states in a two-dimensional disordered potential}.
\newblock Phys. Rev. B \textbf{25}, 2185--2190 (1982).
\newblock \doi{10.1103/PhysRevB.25.2185}.
\newblock \urlprefix\url{https://link.aps.org/doi/10.1103/PhysRevB.25.2185}

\bibitem{haldane:prl1988}
F.D.M. Haldane, {Model for a Quantum Hall Effect without Landau Levels:
  Condensed-Matter Realization of the ``Parity Anomaly''}.
\newblock Phys. Rev. Lett. \textbf{61}, 2015--2018 (1988).
\newblock \doi{10.1103/PhysRevLett.61.2015}.
\newblock \urlprefix\url{https://link.aps.org/doi/10.1103/PhysRevLett.61.2015}

\bibitem{qi:prb2006QWZ}
X.L. Qi, Y.S. Wu, S.C. Zhang, {Topological quantization of the spin Hall effect
  in two-dimensional paramagnetic semiconductors}.
\newblock Phys. Rev. B \textbf{74}, 085308 (2006).
\newblock \doi{10.1103/PhysRevB.74.085308}.
\newblock \urlprefix\url{https://link.aps.org/doi/10.1103/PhysRevB.74.085308}

\bibitem{xiao:prl2022hkspaceBCP}
C.~Xiao, H.~Liu, W.~Wu, H.~Wang, Q.~Niu, S.A. Yang, {Intrinsic Nonlinear
  Electric Spin Generation in Centrosymmetric Magnets}.
\newblock Phys. Rev. Lett. \textbf{129}, 086602 (2022).
\newblock \doi{10.1103/PhysRevLett.129.086602}.
\newblock
  \urlprefix\url{https://link.aps.org/doi/10.1103/PhysRevLett.129.086602}

\bibitem{hoffmann:prb2015}
M.~Hoffmann, J.~Weischenberg, B.~Dup\'e, F.~Freimuth, P.~Ferriani,
  Y.~Mokrousov, S.~Heinze, {Topological orbital magnetization and emergent Hall
  effect of an atomic-scale spin lattice at a surface}.
\newblock Phys. Rev. B \textbf{92}, 020401(R) (2015).
\newblock \doi{10.1103/PhysRevB.92.020401}.
\newblock \urlprefix\url{https://link.aps.org/doi/10.1103/PhysRevB.92.020401}

\bibitem{hanke:prb2016}
J.P. Hanke, F.~Freimuth, A.K. Nandy, H.~Zhang, S.~Bl\"ugel, Y.~Mokrousov, {Role
  of Berry phase theory for describing orbital magnetism: From magnetic
  heterostructures to topological orbital ferromagnets}.
\newblock Phys. Rev. B \textbf{94}, 121114(R) (2016).
\newblock \doi{10.1103/PhysRevB.94.121114}.
\newblock \urlprefix\url{https://link.aps.org/doi/10.1103/PhysRevB.94.121114}

\bibitem{roig:prb2024}
M.~Roig, A.~Kreisel, Y.~Yu, B.M. Andersen, D.F. Agterberg, {Minimal models for
  altermagnetism}.
\newblock Phys. Rev. B \textbf{110}, 144412 (2024).
\newblock \doi{10.1103/PhysRevB.110.144412}.
\newblock \urlprefix\url{https://link.aps.org/doi/10.1103/PhysRevB.110.144412}

\bibitem{solovyev:prb1997}
I.V. Solovyev, {Magneto-optical effect in the weak ferromagnets
  ${\mathrm{LaMO}}_{3}$ (M= Cr, Mn, and Fe)}.
\newblock Phys. Rev. B \textbf{55}, 8060--8063 (1997).
\newblock \doi{10.1103/PhysRevB.55.8060}.
\newblock \urlprefix\url{https://link.aps.org/doi/10.1103/PhysRevB.55.8060}

\bibitem{shao:prapp2021}
D.F. Shao, J.~Ding, G.~Gurung, S.H. Zhang, E.Y. Tsymbal, {Interfacial Crystal
  Hall Effect Reversible by Ferroelectric Polarization}.
\newblock Phys. Rev. Appl. \textbf{15}, 024057 (2021).
\newblock \doi{10.1103/PhysRevApplied.15.024057}.
\newblock
  \urlprefix\url{https://link.aps.org/doi/10.1103/PhysRevApplied.15.024057}

\bibitem{samanta:jappphys2020}
K.~Samanta, M.~Ležaić, M.~Merte, F.~Freimuth, S.~Blügel, Y.~Mokrousov,
  {Crystal Hall and crystal magneto-optical effect in thin films of SrRuO3}.
\newblock Journal of Applied Physics \textbf{127}(21), 213904 (2020).
\newblock \doi{10.1063/5.0005017}.
\newblock \urlprefix\url{https://doi.org/10.1063/5.0005017}.
\newblock
  {\href{https://arxiv.org/abs/https://pubs.aip.org/aip/jap/article-pdf/doi/10.1063/5.0005017/15246940/213904\_1\_online.pdf}{{https://pubs.aip.org/aip/jap/article-pdf/doi/10.1063/5.0005017/15246940/213904\_1\_online.pdf}}}

\bibitem{sivadas:prl2016}
N.~Sivadas, S.~Okamoto, D.~Xiao, {Gate-Controllable Magneto-optic Kerr Effect
  in Layered Collinear Antiferromagnets}.
\newblock Phys. Rev. Lett. \textbf{117}, 267203 (2016).
\newblock \doi{10.1103/PhysRevLett.117.267203}.
\newblock
  \urlprefix\url{https://link.aps.org/doi/10.1103/PhysRevLett.117.267203}

\bibitem{naka:prb2020}
M.~Naka, S.~Hayami, H.~Kusunose, Y.~Yanagi, Y.~Motome, H.~Seo, {Anomalous Hall
  effect in $\ensuremath{\kappa}$-type organic antiferromagnets}.
\newblock Phys. Rev. B \textbf{102}, 075112 (2020).
\newblock \doi{10.1103/PhysRevB.102.075112}.
\newblock \urlprefix\url{https://link.aps.org/doi/10.1103/PhysRevB.102.075112}

\bibitem{naka:prb2021AHEperovskite}
M.~Naka, Y.~Motome, H.~Seo, {Anomalous Hall effect in antiferromagnetic
  perovskites}.
\newblock Phys. Rev. B \textbf{106}, 195149 (2022).
\newblock \doi{10.1103/PhysRevB.106.195149}.
\newblock \urlprefix\url{https://link.aps.org/doi/10.1103/PhysRevB.106.195149}

\bibitem{attias:prb2024}
L.~Attias, A.~Levchenko, M.~Khodas, {Intrinsic anomalous Hall effect in
  altermagnets}.
\newblock Phys. Rev. B \textbf{110}, 094425 (2024).
\newblock \doi{10.1103/PhysRevB.110.094425}.
\newblock \urlprefix\url{https://link.aps.org/doi/10.1103/PhysRevB.110.094425}

\bibitem{chen:prb2022}
H.~Chen, {Electronic chiralization as an indicator of the anomalous Hall effect
  in unconventional magnetic systems}.
\newblock Phys. Rev. B \textbf{106}, 024421 (2022).
\newblock \doi{10.1103/PhysRevB.106.024421}.
\newblock \urlprefix\url{https://link.aps.org/doi/10.1103/PhysRevB.106.024421}

\bibitem{smejkal:natrevmat2022}
L.~\ifmmode~\check{S}\else \v{S}\fi{}mejkal, A.H. MacDonald, J.~Sinova,
  S.~Nakatsuji, T.~Jungwirth, {Anomalous Hall antiferromagnets}.
\newblock Nature Reviews Materials \textbf{7}, 482 (2022).
\newblock \doi{10.1038/s41578-022-00430-3}.
\newblock \urlprefix\url{https://doi.org/10.1038/s41578-022-00430-3}

\end{thebibliography}
\end{document}